\documentclass{article}
\usepackage[preprint]{spconf}
\usepackage{amsmath,graphicx}
\usepackage{booktabs}
\usepackage{multirow, makecell}

\def\ov{\over}

\title{SpecAugment on Large Scale Datasets}
\name{Daniel S. Park\sthanks{Equal contribution.}, Yu Zhang$^*$,
Chung-Cheng Chiu, Youzheng Chen, Bo Li, William Chan,}
\nameplus{Quoc V. Le and Yonghui Wu}
\address{Google Inc.}
\copyrightnotice{\footnotesize\begin{tabular}{p{0.965\textwidth}}\copyright\ 2020 IEEE. Personal use of this material is permitted. Permission from IEEE must be obtained for all other uses, in any current or future media, including reprinting/republishing this material for advertising or promotional purposes, creating new collective works, for resale or redistribution to servers or lists, or reuse of any copyrighted component of this work in other works.\end{tabular}}

\begin{document}

\maketitle

\begin{abstract}
Recently, SpecAugment, an augmentation scheme for automatic speech recognition that acts directly on the spectrogram of input utterances, has shown to be highly effective in enhancing the performance of end-to-end networks on public datasets. In this paper, we demonstrate its effectiveness on tasks with large scale datasets by investigating its application to the Google Multidomain Dataset (Narayanan et al., 2018). We achieve improvement across all test domains by mixing raw training data augmented with SpecAugment and noise-perturbed training data when training the acoustic model. We also introduce a modification of SpecAugment that adapts the time mask size and/or multiplicity depending on the length of the utterance, which can potentially benefit large scale tasks. By using adaptive masking, we are able to further improve the performance of the Listen, Attend and Spell model on LibriSpeech to $2.2\%$ WER on test-clean and $5.2\%$ WER on test-other.
\end{abstract}


\section{Introduction}
\label{sec:intro}

Data augmentation has been a successful method for improving generalization performance in
Automatic Speech Recognition (ASR). Recently, SpecAugment \cite{park2019specaugment},
an augmentation scheme
that directly augments the spectrogram of the input utterance, has shown
surprising effectiveness in improving the performance of ASR networks on the
960h Librispeech and 300h Switchboard datasets.
One natural question that arises is whether the effectiveness of SpecAugment
persists for large scale tasks.

In this paper, we address this question by applying SpecAugment to the Google
Multidomain Dataset introduced in \cite{narayanan2018domaininvariant}.
The Google Multidomain Dataset is a large scale multi-domain dataset,
with multiple test sets from disparate domains.
All data in the data set is anonymized.
We compare the performance of the trained network with
respect to the various forms of augmentation applied to the data, the results
of which are summarized in table \ref{t:summary}.
In \cite{narayanan2018domaininvariant}, Multistyle TRaining (MTR) \cite{kim-interspeech-2017},
where a mixed room simulator is used to combine clean audio with a
large library of noise audio, is employed to augment the input data.
We take this as the baseline when studying the performance of SpecAugment.

\begin{table}[h!]
  \caption{Result of various forms of augmentation.}
  \label{t:summary}
  \centering
  \vskip 0.1in
  \begin{tabular}{ll}
    \toprule
    Data Augmentation & Performance \\
    \midrule
    Multistyle TRaining (MTR) & Baseline \\
    \midrule
    None & Worse \\
    SpecAugment & Better$^*$ \\
    SpecAugment + MTR & Worse \\
    Mix SpecAugment \& MTR & Better \\
    \bottomrule
  \end{tabular}
\end{table}

As summarized in table \ref{t:summary}, we compare the performance
of the network when trained on clean data, data with MTR applied,
data with SpecAugment applied, data with both SpecAugment and
MTR applied, and data obtained by mixing SpecAugmented
and MTR data. We find that SpecAugment, when applied to clean data,
performs better than the baseline on all natural test sets,
while it performs worse only on a synthetic test set obtained by
applying MTR to test utterances.
To our surprise, applying SpecAugment on top of
MTR degrades performance across most domains.
Meanwhile, we are able to achieve improvement across all domains
by mixing SpecAugmented data with MTR data.

SpecAugment requires a negligible amount of additional computational
resources, does not require additional audio data, can be applied online
and is thus highly scalable as the training set becomes large.
Our results therefore suggest that SpecAugment can be considered as
a serious alternative to more sophisticated resource-heavy augmentation
methods.

SpecAugment policies consist of frequency masking, time masking and time warping.
The augmentation policies considered in \cite{park2019specaugment} have a fixed
number of time masks regardless the length of the utterance.
On large scale tasks spanning multiple domains,
we expect the length of the utterances to have a large variance.
We thus introduce adaptive time masking, where
the number of time masks and/or the size of the time mask
vary depending on the length of the input.
We experiment with several adaptive policies on the Google Multidomain Dataset
and LibriSpeech 960h \cite{Panayotov2015LibriSpeechAA}.
So far, we have not found adaptive time policies that perform better than
vanilla SpecAugment on the Google Multidomain Dataset.
Meanwhile, we find adaptive policies that yield performance gains
on LibriSpeech relative to \cite{park2019specaugment},
as we are able to train a Listen, Attend and Spell \cite{chan2015listen} network
to have $2.2\%$ WER on test-clean and $5.2\%$ WER on test-other.

\subsection{Related Work}

There is a vast literature on augmentation in ASR, only a part of which we survey here.
Artificial data augmentation for low resource speech recognition tasks has been
studied in \cite{kanda-asru-2013,ragni-interspeech-2014}.
Vocal Tract Length Perturbation has been introduced in the context of data augmentation
for ASR in \cite{jaitly-2013-icml}, and explored further in \cite{kim2019improved}.
Noisy audio signals have been used for augmentation in \cite{hannun-arxiv-2014}.
Speed perturbation \cite{ko-interspeech-2015} has been an integral part of augmentation
of speech data. The work \cite{kim-interspeech-2017} studies the effect of using
acoustic room simulators. The works \cite{prabhavalkar-2015-icassp, raju-arxiv-2018}
examine application of data augmentation for keyword spotting.
Drop-out for features have been used for training multi-stream ASR systems
in \cite{mallidi-icassp-2016, kovacs2019examining}.
Systematic omission of frequency channels of the input spectrogram
has been studied in the context of CNN ASR networks in
\cite{kovacs2017increasing,toth2018perceptually}.
We have commented on SpecAugment \cite{park2019specaugment}
in the introduction.

Data augmentation has also been successfully applied to large scale industrial datasets. As noted earlier, Multistyle TRaining (MTR) is a popular technique where clean audio is combined with background noise using a room simulator \cite{kim-interspeech-2017}. MTR has been successfully applied to HMM-based systems \cite{sainath-interspeech-2015, li-interspeech-2017} and end-to-end LAS models \cite{chan2015listen, chiu-icassp-2018, chiu-interspeech-2018}. A natural question is how SpecAugment compares to or can complement existing data augmentation techniques like MTR, especially on large scale datasets.

Our contribution in this paper is three-fold:
\begin{enumerate}
    \item We scale up SpecAugment to large scale industrial datasets. We compare to existing MTR data augmentation, and present how we can improve upon it.
    \item We demonstrate that SpecAugment improves the performance of streaming models.
    \item We present an adaptive version of SpecAugment, where the degree of time masking is adaptive to the input sequence length.
\end{enumerate}

\section{SpecAugment and Adaptive Masking}

We briefly review SpecAugment in this section,
and introduce its adaptive variants. A SpecAugment policy is obtained
by composing three basic augmentations---time warping, frequency masking
and time masking. We denote the time and frequency dimensions of
the spectrogram as $\tau$ and $\nu$.
\begin{enumerate}
\item Time warping with parameter $W$: A displacement $w$ is chosen from a uniform
distribution from $-W$ to $W$. A start point $w_0$ is chosen from the time interval
$[W,\tau-W)$. A linear warping function $\mathcal{W}(t)$ is defined so that the
start point $w_0$ is mapped to the point $w_0 + w$ and that the boundary points
$t=0$ and $t=\tau-1$ are fixed:
\begin{equation}
\mathcal{W}(t) = 
\begin{cases}
\left( {w_0 + w \ov w_0} \right) t & t \leq w_0 \,, \\
{(\tau - 1 - w_0 - w)t + (\tau - 1)w \ov \tau - 1 - w_0} & t > w_0 \,.
\nonumber
\end{cases}
\end{equation}
Warping is defined so that the warped features $\mathbf{x}_\text{warp}(t)$
(in our case, log-mel frequency coefficients) at time $t$ are related to the
original features $\mathbf{x}_\text{orig}(t)$ by
\begin{equation}
\mathbf{x}_\text{warp} (\mathcal{W}(t)) = \mathbf{x}_\text{orig}(t)\,.
\nonumber
\end{equation}
We note that the original implementation of time warping presented in
\cite{park2019specaugment}, for all practical purposes, is equivalent to
this alternative definition.
\item Frequency masking with parameter $F$: A mask size $f$ is chosen
from a uniform distribution from 0 to $F$. The consecutive log-mel frequency channels
$[f_0, f_0 + f)$ are then masked, where $f_0$ is chosen from $[0, \nu -f)$.
\item Time masking with parameter $T$: A mask size $t$ is chosen from a uniform
distribution from 0 to $T$. The consecutive time steps $[t_0, t_0 + t)$
are masked, where $t_0$ is chosen from $[0, \tau-t)$.
\end{enumerate}
The SpecAugment policies in \cite{park2019specaugment} consist of applying
these three augmentations a fixed number of times.

In large scale datasets that contain disparate domains of inputs,
we expect there to be a large variance in the length of the input audio.
Thus, a fixed number of time masks may not be adequate for such tasks,
as the time masking may be too weak for longer utterances, or too severe for
shorter ones. We thus introduce two different
ways time masking can be made adaptive with respect to length of the spectrogram $\tau$:
\begin{enumerate}
\item Adaptive multiplicity: The number, or multiplicity, of time masks
$M_\text{t-mask}$ is set to be
$M_\text{t-mask} = \lfloor p_M \cdot \tau \rfloor$ for the multiplicity ratio $p_M$.
\item Adaptive size: The time mask parameter is set to be
$T = \lfloor p_S \cdot \tau \rfloor$ for the size ratio $p_S$.
\end{enumerate}
In this paper, we cap the number of time masks at 20 when using adaptive time
masking, so that $M_\text{t-mask}$ is given by
\begin{equation}
\nonumber
M_\text{t-mask} = \text{min} (20, \lfloor p_M \cdot \tau \rfloor) \,.
\end{equation}

\section{Experiments}

\subsection{LibriSpeech 960h}

\subsubsection{Set-up}

Our set-up for LibriSpeech 960h is based on that of \cite{park2019specaugment}.
We use the model LAS-6-1280 of that work and train with training schedule ``L"(ong).
We use shallow fusion \cite{gulcehre-2015-arxiv} with an LSTM language model (LM)
with two fusion parameters---the LM weight and coverage penalty
\cite{chorowski-2017-interspeech}. In this work, we use a 3-layer LSTM with width 4096,
with a resulting word-level perplexity of 63.6 on the dev-set transcripts.
We tune the fusion parameters on the dev-set
using grid-search and apply them to the test set to report the final results.

\subsubsection{Adaptive SpecAugment Policies}

We compare three augmentation policies. The baseline policy is the policy coined
``LibriSpeech Double" in \cite{park2019specaugment}. This policy has two frequency
masks with $F =27$, two time masks with $T=100$ which are applied after
time warping with $W = 80$.

Let us introduce a hand-crafted adaptive policy, which we denote \textbf{LibriFullAdapt}.
This policy has two frequency mask applications with $F=27$
and time masks with both adaptive multiplicity and size with $p_M=0.04$ and $p_S=0.04$
applied on top of time warping applied with $W=80$.

\subsubsection{Results}

We list the results of our training in table \ref{t:libriSOTA}.
We find that the adaptive policy performs better than
the fixed policy, and observe gain in performance both before and
after shallow fusion with the language model.

\begin{table}[h!]
  \caption{LibriSpeech 960h WERs (\%).}
  \label{t:libriSOTA}
  \vskip 0.1in
  \centering
  \small
  \resizebox{\columnwidth}{!}{%
  \begin{tabular}{lcccc}
    \toprule
    \bfseries Method & \multicolumn{2}{c}{\bfseries No LM} & \multicolumn{2}{c}{\bfseries With LM} \\
    \cmidrule(r){2-3} \cmidrule(r){4-5}
     & \bfseries clean & \bfseries other & \bfseries clean & \bfseries other \\
    \midrule
    \bfseries Recent Work \\
    \quad SpecAugment \cite{park2019specaugment} & 2.8 & 6.8 & 2.5 & 5.8 \\
    \quad L\"uscher et al., (2019) \cite{luscher2019rwth} & & & 2.3 & \bfseries 5.0 \\
    \quad Kim et al., (2019) \cite{kim2019improved} & & & 2.4 & 8.3 \\
    \quad Karita et al., (2019) \cite{karita2019comparative} & & & 2.6 & 5.7 \\
    \quad Han et al., (2019) \cite{han2019stateoftheart} & & & \bfseries 2.2 & 5.8 \\
    \midrule
    \bfseries This Work \\
    \quad LAS + Baseline SpecAugment & 2.8 & 6.8 & 2.4 & 5.7 \\
    \quad LAS + LibriFullAdapt & \bfseries 2.6 & \bfseries 6.0 & \bfseries 2.2 & 5.2 \\
    \bottomrule
  \end{tabular}
  }
  \vskip -0.1in
\end{table}

\subsection{Google Multidomain Dataset}

\subsubsection{Data and Augmentation}

We study the effect of SpecAugment when training on the Google Multidomain
Dataset \cite{narayanan2018domaininvariant}. We consider five test
sets---Search, Search-Noisy, TTS-Audiobook, Telephony and YouTube---to
measure the performance of the network. All training and testing data
is anonymized.

\begin{table*}[t!]
\caption{Performance of various augmentation schemes on the Google Multidomain Dataset.}
\label{t:main_table}
\centering
\small
\begin{tabular}{lccccc}\\\toprule 
\textbf{Augmentation} & \textbf{Search} & \textbf{Search-Noisy} & \textbf{TTS-Audiobook} & \textbf{Telephone} & \textbf{YouTube} \\\midrule
MTR (Baseline) & 6.6 & 9.6 & 4.6 & 7.8 &  11.8 \\
\midrule
\bfseries Control Experiments \\
\quad Clean Data & 6.7 & 14.3 & 4.5 & 10.3 & 11.3\\
\quad SpecAugBasic + MTR & 6.9 & 9.7 & 4.5 & 8.2 & 10.8\\
\quad Frequency Masking Only & 6.4 & 13.4 & 4.8 & 8.0 & 11.4 \\
\midrule
\bfseries SpecAugment \\
\quad SpecAugBasic on Clean & \bf{6.2} & 12.9 & \bf{4.2} & \bf{7.2} & \bf{10.3}\\
\quad SpecAugBasic \& MTR (20\%) Mixed & 6.3 & \bf{9.4} & \bf{4.2} & \bf{7.2}& 10.4\\
\bottomrule
\end{tabular}
\end{table*}

As a baseline for our experiments, we augment the input data by using a room
simulator described in \cite{kim-interspeech-2017}. For training,
various factors of the room simulator, including room-size, reverberation time,
microphone positions, speech and noise sources, signal to noise ratio are
randomly selected and applied to all input utterances. The injected noise
is sampled from either anonymized YouTube audio or a
collection of real-life noises.
The test set Search-Noisy is constructed by applying these perturbations
to the Search test set.

The network input is a log-mel frequency spectrogram obtained from the audio
using 32 msec frame windows with 10 msec shift.
The log-mel frequency coefficients have 128 dimensions, and are stacked with height
512 with stride 3. The text is tokenized using a Word Piece Model (WPM)
\cite{schuster-icassp-2012} of vocabulary size 4k.

We consider five different input configurations: MTR data, clean data,
MTR data with SpecAugment applied, clean data with SpecAugment applied
and finally data obtained by mixing clean data with SpecAugment applied
and MTR data with an 8:2 ratio.
Augmentation is applied to the spectrogram after
unstacking the features to obtain an array of 128 dimensional features.
The augmented spectrogram is then restacked to the original form and fed
into the acoustic model.

We present the result of training with a vanilla SpecAugment policy, which
we denote \textbf{SpecAugBasic}. This policy has two frequency masks
and two time masks with $T=50$. Time warping has not been used.
As a control experiment, we also train the network on data augmented
only using frequency masking with two masks of $F=27$.

\subsubsection{SpecAugmemt on RNN Transducer (RNN-T)}

We train an RNN-T model described in \cite{he2018streaming}.
The encoder is an 8-layer uni-directional LSTM with cell size 2048, while the decoder is a 2-layer LSTM with the same cell size. No language model is used.

We note that this model produces weaker context information due to its streaming nature. We nevertheless get gains from time masking, as we demonstrate shortly.

As explained in \cite{he2018streaming}, our RNN-T model heavily relies on layer normalization \cite{ba2016layer}.
Note that the application of time masks make the variance of hidden activations vanish, which destabilizes training in the presence of layer normalization. Even when using an aggressive variance floor, this still leads to huge gradients when the network becomes deeper. To alleviate this instability, we add Gaussian noise to the time masked regions, which stabilizes training.

\subsubsection{Results}

The results of training the acoustic model using the different augmentation
methods are presented in table \ref{t:main_table}. Note that when SpecAugment
is applied on top of MTR, the performance degrades below the baseline
across all test sets.

Meanwhile, we find that when SpecAugBasic is applied to the clean utterances, it
out-performs the baseline across all ``natural test sets," while it performs worse on
the synthetic test set obtained by applying MTR to Search-domain utterances.
This degradation, however, can be addressed by ensembling SpecAugmented data
with MTR data, as shown in the last row of the table.

We note that while we have experimented with adaptive time masking
policies, we have not discovered one that out-performs fixed policy
SpecAugBasic. The benefit of adaptive time masking on this dataset
has yet to be seen.

We emphasize that the trained model is a streaming model,
whose performance SpecAugment is still able to noticeably improve.
Furthermore, we see that time masking plays an important role in
improving the performance of this network, which is evident from
the evaluation results on the YouTube dataset.

\section{Summary and Discussion}

We find that SpecAugment, despite its simplicity, yields better gains
on large scale datasets compared to time-tested and more sophisticated
augmentation methods. Given the computational advantage that SpecAugment has,
we find it has rich potential for being incorporated into the
data pipeline of industrial-scale tasks.

We have introduced adaptive time-masking for SpecAugment.
While we have not been able to find an adaptive policy that out-performs
a non-adaptive policy on the Google Multidomain Dataset,
we have demonstrated the effectiveness of adaptive masking
on LibriSpeech 960h. We expect further exploration
of adaptive masking to bring improvements when SpecAugment
is applied to large scale tasks.

\section{Acknowledgement}

We thank Yuan Cao, Ekin Dogus Cubuk, Yanping Huang, Luke Metz, Arun Narayanan, Ruoming Pang, Tara Sainath, Qizhe Xie and Barret Zoph for useful discussions and helping with our experiments.

\bibliographystyle{IEEEbib}
\bibliography{strings,refs}

\end{document}